# Dynamics of entrapped microbubbles with multiple openings

Amit Dolev*, Murat Kaynak, Mahmut Selman Sakar

Institute of Mechanical Engineering, École Polytechnique Fédérale de Lausanne, CH-1015 Lausanne, Switzerland

*Corresponding author

| | |
|---|---|
| E-mail address: | amitdtechnion@gmail.com |
| Postal address: | MED 3 2715 (Bâtiment MED) Station 9 CH-1015 Lausanne, Switzerland |

Author e-mail addresses:

Murat Kaynak:           murat.kaynak@epfl.ch

Mahmut Selman Sakar:    selman.sakar@epfl.ch

## ABSTRACT

Microbubbles excited by acoustic fields inside water oscillate, and generate acoustic radiation forces and drag-induced acoustic streaming. These forces can be harnessed in various biomedical applications such as targeted drug delivery and on-chip biomanipulation. The conventional approach for using microbubbles as actuators is to trap them inside microfabricated cavities. Anisotropic forces are applied by constraining the interfaces where the air interacts with water . The existing analytical models derived for spherical bubbles are incapable of predicting the dynamics of bubbles in such configurations. Here, a new model for bubbles entrapped inside arbitrarily shaped cavities with multiple circular openings is developed. The semi-analytical model captures a more realistic geometry through a solution to an optimization problem. We challenge the assumption that bubbles should be excited at their first resonance frequency to optimize their performance. The natural frequencies and the correlated normal vibration modes are calculated, which are subsequently used to compute the acoustic streaming patterns and the associated thrust by a finite element simulation. An experimental platform was built to measure the deflection of beams loaded by microfabricated bubble actuators, and visualize the generated streaming patterns. The results highlight the contribution of the computational model as a design tool for engineering applications.

# 1. Introduction

Microbubbles trapped inside microcavities engraved on the walls of microfluidic devices have been used as wireless actuators for on-chip manipulation of particles, cells, and entire organisms [1,2]. Likewise, microbubbles encapsulated within untethered structures serve as propellers for mobile micromachines [3–5]. In an acoustic field, entrapped microbubbles oscillate and generate two different types of forces: drag-induced microstreaming (i.e., confined mean flow [6,7]) and radiation forces [8]. It has been widely accepted that microbubbles deliver the best performance when the acoustic field is tuned to the resonance frequency of the microbubble [3,4,9,10]. As a manifestation of this assumption, experimental studies reported the frequency at which the velocity of the surrounding fluid or bubble-propelled machine peaked as the resonance frequency of the microbubble. However, the reported results may be misleading because the properties of the driving acoustic signal (e.g., the pressure generated by the transducer) have been ignored or not fully characterized. Futheremore, directly measuring the displacement at the surface of the microbubble at different excitation frequencies is very challenging, particularly when the bubble is not stationary, because the bubble deformation occurs in 3D and the vibration amplitude is up to a few microns [5,11,12]. Optimizing the performance of microbubble actuators is essential for the development of next-generation biomedical devices and untethered microrobots [13,14].

A complete understanding of microbubble dynamics involves the calculation of vibration modes, corresponding natural frequencies, and generated thrust due to acoustic streaming (AS). Theoretical work has been focused on basic configurations—free bubbles with a spherical geometry [8,15]. The formulation involves an analytical approximation for the natural frequencies and associated vibration modes. Notable examples are models of gas bubbles in inviscid fluids [15,16], in viscoelastic fluids [17], or encapsulated by a viscoelastic shell [18]. The case of entrapped bubbles with a single opening or multiple openings attracted less attention. The early work by Miller and Nyborg provides an approximation for the first resonance frequency of a cylindrical gas-filled pore on an infinite surface [19]. The governing equation was derived by assuming a parabolic vibration mode. Gelderblom *et al.*, extended the model presented in [19] to include axisymmetric vibration modes by solving fluid-gas coupled problems [20]. The gas was modeled as an ideal gas while the fluid was modeled using potential flow in the lossless case and unsteady stokes flow in the lossy case. Gritsenko *et al.* studied the natural frequencies and general vibration modes (i.e., axisymmetric and non-axisymmetric) of baffled cylindrical gas-filled pores assuming the gas-fluid interface was clamped [21]. Notably, the coupled gas and fluid fields were modeled differently, using velocity potentials, following the work by Chindam *et al.*, [22]. Schnitzer *et al.*, systematically studied bubbles trapped in microgrooves, considering their resonance frequencies and acoustic interactions [23]. The gas and fluid fields were modeled using acoustic equations, namely, Helmholtz equations, because the primary objective was to study acoustic phenomena. Spelman *et al.*, studied entrapped spherical bubbles with multiple openings [10]. Their formulation was based on the analysis of fluid immersed spherical gas bubbles with constraints [24].

Radiation forces and AS generated by vibrating bubbles are nonlinear phenomena that can seldom be formulated and solved analytically. Therefore, they have been calculated numerically by employing various methods including finite element method (FEM), boundary element method (BEM), and computational fluid dynamics (CFD). The research in this domain has been focused on the primary forces acting on particles and bubbles in acoustic fields, and not on the secondary forces generated by the excited structures. These secondary effects give rise to additional radiation forces and AS, which are proven to be effective in micromanipulation. Although knowing the streaming patterns and thrust generated by the secondary fields is important, to our knowledge, they have not been correlated to the vibration modes of entrapped microbubbles. The primary forces which are due to the primary fields can be computed for various geometries using different methods. Dolev *et al.*, computed the acoustic field using BEM and then employed Gor'kov's theory to estimate the acoustic radiation forces acting on an acoustically levitated rigid sphere [25]. Muller *et al.*, provided FEM based methodology to calculate both the acoustic radiation force and AS induced drag force acting on small particles [26]. Their methodology was built upon the perturbation solution of the thermoacoustic equations. Behdani *et al.* studied the streaming patterns generated by the oscillation of a microbubble in a microfluidic channel by a direct numerical simulation – CFD [27].

In this paper, we study the acoustic excitation of entrapped gas microbubbles with multiple openings, the streaming patterns, and generated thrust. We focus on natural frequencies and not resonance frequencies for three reasons. First of all, radiation forces and AS are acoustic phenomena that are driven by the gas-fluid interface velocity while the maximum velocities are obtained at the natural frequencies [28]. Second, the natural frequencies and normal modes are independent of the damping in the systems and, thus, can be treated as inherent characteristics of the linear system. Lastly, as demonstrated by Gelderblom *et al.*, [20], resonance frequencies of a gas pocket can be reliably predicted by the potential flow model (i.e., lossless). Regarding the last point, when using modal analysis, the relation between the resonance frequency and natural frequency is

$$\omega_r = \omega_n \sqrt{1 - 2\zeta^2} \,, \tag{1}$$

where $\omega_r$ is the resonance frequency, $\omega_n$ is the natural frequency, and $\zeta$ is the modal damping. Following Eq.(1), $\omega_r \approx \omega_n$ when the damping is low (i.e., almost lossless). To solve the natural frequencies and vibration modes for the general case, which involves multiple openings and a finite surface, a semi-analytical method is proposed. The method is based on an optimization problem that couples the analytical solutions with a BEM simulation. Once the natural frequencies and vibration modes are estimated, the solution is used in a FEM simulation to compute the streaming patterns and generated thrust [26,29]. The numerical results showed that different vibration modes generate different streaming patterns and that the thrust is highly dependent on the excitation frequency. On one hand, higher frequencies lead to stronger thrust, but on the other hand, higher modes generate weaker thrust.

To verify the assumptions of the model and validate the simulation results, experiments were carried out using a platform comprising a water tank, a hydrophone, an imaging system, and a computer-controlled ultrasonic transducer. The tank was placed on top of an inverted microscope to which a high-speed camera was attached. We performed experiments on microfabricated polymer devices encapsulating microbubbles with two openings. The high-speed camera recorded the streaming patterns and the deflection of the structures. The results show that different modes were excited.

The rest of the paper is organized as follows. In Section 2, we derive the governing equations of a trapped bubble with multiple openings. In Section 3, we introduce the coupled optimization problem. FEM simulations for the computation of the streaming patterns and thrust are described in Section 4. The same section contains a sensitivity analysis for the model parameters. In Section 5, a microdevice entrapping a bubble with two openings is studied. Its governing equations of motion including damping and external forcing are provided. The theoretical analysis is verified experimentally, demonstrating the ability to excite different modes. Finally, we present the conclusions from the study in Section 6.

## 2. Natural frequencies and vibration modes – extreme cases

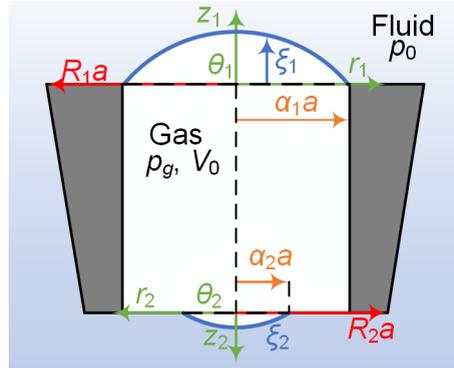

**Fig. 1.** The geometry of the problem. A gas-filled cavity with two interfaces immersed in a fluid. The solid walls entrapping the bubble are denoted with black contour, the solid is colored grey, and the gas-fluid interfaces are highlighted with blue lines. The local coordinate systems are sketched in green, and the dimensions are given in orange and red.

*Governing equations of motion*

The governing equations for the lossless case of a fluid-immersed, entrapped gas bubble with multiple openings are derived by following the work of Gelderblom *et al.*[20]. The gas is assumed to be ideal, and the fluid domain is modeled assuming potential flow. To simplify the math, we derive the equations for a cavity with two interfaces as shown in Fig. 1.

The complete list of modeling assumptions is as follows:

1) The cavity has circular openings.
2) The interface is pinned to the circular edge of the cavity.
3) The surrounding fluid is incompressible and inviscid.
4) The interface is flat at the equilibrium.
5) Lossless problem.
6) The gas flow within the cavity is negligible.
7) Linear harmonic analysis.
8) Polytropic process in the gas.
9) There is no interaction between the interfaces through the liquid.
10) The acoustic wavelength in the cavity is much larger than the size of the cavity.

The local deflection of each interface is described by $\xi_i(r_i, \theta_i, t)$, $i = 1,2$, with a local cylindrical coordinate system located on the axis of each cavity's opening. Employing lossless harmonic analysis, the interfaces can be described as:

$$\xi_i(r_i, \theta_i, t) = \xi_i(r_i, \theta_i) e^{j\omega t}, \quad j = \sqrt{-1}. \tag{2}$$

Each interface motion is coupled to the velocity field in the liquid through a kinematic condition that yields:

$$u_{iz} = j\omega \xi_i, \tag{3}$$

here, $\mathbf{u}_i = u_{ir}\hat{\mathbf{r}} + u_{iz}\hat{\mathbf{z}}$ is the velocity field in the fluid near the opening $i$.

The boundary conditions for the fluid domain are as follows:

On the interface, due to the 4th assumptions, and the kinematic condition:

$$u_{iz}\big|_{z_i=0} = j\omega \xi_i, \quad 0 \leq r_i/a \leq \alpha_i. \tag{4}$$

On the solid surface ($z_i$ =0), no penetration and no-slip conditions are denoted as:

$$u_{iz}\big|_{z_i=0} = 0, \quad \alpha_i \leq r_i/a \leq R_i, \tag{5}$$

$$u_{ir}\big|_{z_i=0} = 0, \quad \alpha_i \leq r_i/a \leq R_i. \tag{6}$$

We ignore the boundary conditions on the cavity walls. The coupling between the pressures in the liquid and the gas occurs via the dynamic boundary conditions at $z_i$=0.

$$p_g = p_i + \sigma C_i - 2\mu \frac{\partial u_{iz}}{\partial z_i}\bigg|_{z_i=0}. \tag{7}$$

Here, $p_g$ is the pressure in the gas bubble, $p_i$ is the pressure in the liquid adjacent to the $i$th interface, $\sigma$ is the surface tension coefficient, $C_i$ is the curvature of the free surfaces $\xi_i$, and $\mu$ is the fluid dynamic viscosity. As a result of the 3rd assumption, $\mu$ is neglected. Considering small deflections, the curvature for the free surfaces is approximated by

$$C_i \approx -\left(\frac{1}{r_i}\frac{\partial \xi_i}{\partial r_i} + \frac{\partial^2 \xi_i}{\partial r_i^2} + \frac{1}{r_i^2}\frac{\partial^2 \xi_i}{\partial \theta_i^2}\right). \tag{8}$$

We assume a general polytropic relation between the instantaneous gas volume $V$ and the gas pressure $p_g$. Expanding the relation for small volume variations yields

$$p_g = p_0 \left(\frac{V_0}{V}\right)^\kappa \approx p_0 \left[1 - \kappa \left(\frac{V}{V_0} - 1\right)\right], \tag{9}$$

where $V_0$ is the cavity's volume when the interfaces are flat, $\kappa$ is the gas polytropic index, and $\kappa=1$ is applicable for isothermal conditions. The instantaneous gas volume can be found from the shape of the interfaces by the following integration:

$$V = V_0 \left(1 + \sum_i \frac{a^3}{V_0} \frac{\int_0^{2\pi} \int_0^{a\alpha_i} \xi_i(r_i, \theta_i, t) r_i dr_i d\theta_i}{a^3}\right) = V_0 \left(1 + \lambda \sum_i \frac{H_i}{a^3}\right),$$

$$H_i = \int_0^{2\pi} \int_0^{a\alpha_i} \xi_i(r_i, \theta_i, t) r_i dr_i d\theta_i, \quad \lambda = \frac{a^3}{V_0}. \tag{10}$$

Substituting Eqs.(9)-(10) to Eq. (7), the dynamic boundary condition becomes

$$p_0 \left(1 - \kappa\lambda \sum_i \frac{H_i}{a^3}\right) = p_i - \sigma \left(\frac{1}{r_i} \frac{\partial \xi_i}{\partial r_i} + \frac{\partial^2 \xi_i}{\partial r_i^2} + \frac{1}{r_i^2} \frac{\partial^2 \xi_i}{\partial \theta_i^2}\right). \tag{11}$$

From Eq. (11) and the first two assumptions, it is intuitive to select the vibration modes of a circular membrane as basis functions for spanning the vibration modes.

According to the 3rd assumption, the flow is irrotational, and the velocity field can be written in terms of a velocity potential. Moreover, due to the 9th assumption, the velocity fields near each opening are assumed to be local, and thereby can be described by local potentials as follows:

$$\mathbf{u}_i = \nabla \phi_i. \tag{12}$$

The following dimensionless quantities are used from here forth without the hat symbol $\hat{\bullet}$.

$$\hat{r}_i = \frac{r_i}{a}, \quad \hat{z}_i = \frac{z_i}{a}, \quad \hat{\xi}_i = \frac{\xi_i}{a}, \quad \hat{u}_i = \sqrt{\frac{a\rho}{\sigma}} u_i, \quad \hat{\phi}_i = \sqrt{\frac{\rho}{a\sigma}} \phi, \quad \hat{t} = \sqrt{\frac{\sigma}{a^3 \rho}} t, \quad \hat{p}_i = \frac{a}{\sigma} p_i, \quad \hat{H}_i = \frac{H_i}{a^3}. \tag{13}$$

*Potential and kinetic energies*

The system is considered lossless and Lagrange equations can be employed to derive the governing equations.

$$\frac{d}{dt}\left(\frac{\partial \mathcal{L}}{\partial \dot{q}_l}\right) - \frac{\partial \mathcal{L}}{\partial q_l} = 0, \quad l = 1, 2, \ldots, \quad \mathcal{L} = \mathcal{E}_k - \mathcal{E}_p \tag{14}$$

where the overdot • indicates a derivative with respect to time *t*, **q** is a set of generalized degrees of freedom (DOF), $\mathcal{L}$, $\mathcal{E}_k$ and $\mathcal{E}_p$ are the Lagrangian, kinetic energy, and potential energy of the system, respectively. The contribution of the irrotational flow near each interface to the kinetic energy is expressed as [30]:

$$\mathcal{E}_{ki} = \frac{1}{2}\int_A \phi_i \frac{\partial \phi_i}{\partial \hat{\mathbf{n}}} dA, \qquad (15)$$

where $\hat{\mathbf{n}}$ is a unit vector pointing out of the fluid. Far from the interfaces, the velocity reduces to zero. The impermeability conditions imply that the total kinetic energy can be reduced to

$$\mathcal{E}_k = \sum_i \mathcal{E}_{ki}, \quad \mathcal{E}_{ki} = -\frac{1}{2}\int_0^{2\pi}\int_0^{\alpha_i} \phi_i\big|_{z_i=0} \frac{\partial \xi_i}{\partial t} r_i dr_i d\theta_i. \qquad (16)$$

The potential energy of the system is related to the surface tension and the volume of the microbubble. The energy increases as the area of the interface grows and the bubble volume decreases.

$$\mathcal{E}_p = \sum_i \int_{A_{i0}}^{A_i} dA_i - \int_{V_0}^{V}(p_g - p_0)dV \qquad (17)$$

Substituting Eq.(9) and assuming small amplitude variations, the potential energy is approximated as

$$\mathcal{E}_p \approx \frac{1}{2}\sum_i \int_0^{2\pi}\int_0^{\alpha_i} \frac{1}{r_i}\left(\frac{\partial \xi_i}{\partial \theta_i}\right)^2 + r_i\left(\frac{\partial \xi_i}{\partial r_i}\right)^2 dr_i d\theta_i + \frac{\kappa \lambda p_0}{2}\sum_i\sum_j H_i H_j \qquad (18)$$

To solve the general problem for which $0 < R_i \leq \infty$, we first solve for two extreme cases, free ($R_i$=0) and baffled ($R_i$=∞). Notice that the free case is not physical.

### 2.1.1. The free bubble

To solve the interface displacement fields, we span them using a set of basis functions. As mentioned in Section 2.1, these functions are essentially the vibration modes of a circular membrane [31].

$$\xi_i(r_i,\theta_i,t) = \sum_{m=0}^{\infty}\sum_{n=1}^{\infty} J_m\left(j_{mn}\frac{r_i}{\alpha_i}\right)\left[q_{imnA}\cos(m\theta_i) + q_{imnB}\sin(m\theta_i)\right], \quad m = 0,1,\ldots \quad n = 1,2,\ldots \qquad (19)$$

Here, $J_m$ is a Bessel function of the first kind of order $m$, $j_{mn}$ is the $n^{\text{th}}$ zero of $J_m$, and $q_{imnA}$ and $q_{imnB}$ are time-dependent unknown functions. Because the potential energy depends only on the gas bubble, substituting Eq.(19) into Eq. (18) yields Eq. (A1). In the free bubble case, the only boundary condition to be satisfied is given by Eq.(4), and the following potential fields are assumed [21]

$$\phi_i^{(F)}(r_i,\theta_i,z_i,t) = -\sum_{m=0}^{\infty}\sum_{n=1}^{\infty}\frac{1}{j_{mn}}J_m\left(j_{mn}\frac{r_i}{\alpha_i}\right)\left[\dot{a}_{imn}\cos(m\theta_i)+\dot{b}_{imn}\sin(m\theta_i)\right]e^{-j_{mn}z_i}, \quad (20)$$

where $a_{imn}$ and $b_{imn}$ are time-dependent unknown functions. To compute $a_{imn}$ and $b_{imn}$, the derivative of Eq.(20) with respect to $z_i$ is computed. By using the boundary condition and the orthogonality properties of the trigonometric and Bessel functions, they are obtained as Eq.(A2). Substituting Eq.(A2). into Eq.(20), and the kinetic energy for a free bubble independent of $a_{imn}$ and $b_{imn}$ is obtained as Eq.(A3).

### 2.1.2. The baffled bubble

We assume that the interface displacement fields can be spanned by the same basis functions as for the free bubble, and the boundary conditions to be satisfied are given by Eqs.(4) and (5). To satisfy them, the following potential fields are assumed [20]

$$\phi_i^{(B)}(r_i,\theta_i,z_i,t) = -\int_0^{\infty}\sum_{m=0}^{\infty}\Phi_{im}(k)J_m\left(k\frac{r_i}{\alpha_i}\right)\left[\dot{a}_{imn}\cos(m\theta_i)+\dot{b}_{imn}\sin(m\theta_i)\right]e^{-kz_i}dk. \quad (21)$$

To compute $a_{imn}$ and $b_{imn}$ the derivative of Eq.(21) with respect to $z_i$ is computed, then by using the boundary conditions and the orthogonality properties of the trigonometric and Bessel functions, they are obtained as Eq.(A4). Then, the kinetic energy for a baffled bubble is obtained as Eq.(A5).

### 2.1.3. The intermediate case

The intermediate case, where $R<\infty$, is a realistic situation that is more relevant to physical scenarios than the free and baffled cases. The potentials should comply with the boundary conditions given by Eqs.(4) and (6). However, no analytical potential satisfying these two conditions was found. We assume that the solution of the intermediate case lies in between the two extreme ones and introduced the following potential:

$$\phi_i^{(I)} = e^{-\beta_i(R_i-\alpha_i)}\phi_i^{(F)} + \left(1-e^{-\beta_i(R_i-\alpha_i)}\right)\phi_i^{(B)} \quad (22)$$

where $\beta_i$ are unknown coefficients, which are found by solving an optimization problem as described in Section 3.

*Discrete linear equations of motion*

The problem is studied in the framework of linear harmonic analysis. It follows that every arbitrary time-dependent function can be described as $F(t) = Fe^{j\omega t}$. Notice that the frequency $\omega$ is normalized according to Eq.(13). Now, to discretize the continuous equations, the coefficient $q_{imnA}$ and $q_{imnB}$ are used as discrete DOF, indicating the participation factor of each basis function. Because the energies hold terms up to the second-order, utilizing the Lagrange equations yields discrete linear equations of motion. Similarly, we can directly derive the mass, **M**, and stiffness, **K**, matrices, and write the equations of motion as follows:

$$\mathbf{M}\ddot{\mathbf{q}} + \mathbf{K}\mathbf{q} = \mathbf{0} \tag{23}$$

The DOF in **q** can be organized differently, here, we use a finite set of cosine DOF (i.e., $q_{imnA}$), and order them as follows:

$$\mathbf{q} = \{q_{101A} \quad q_{102A} \quad \cdots \quad q_{10NA} \quad q_{111A} \quad q_{112A} \quad \cdots \quad q_{11NA} \quad \cdots \quad q_{1M1A} \quad q_{1M2A} \quad \cdots \quad q_{1MNA} \quad \cdots$$
$$q_{201A} \quad q_{202A} \quad \cdots \quad q_{20NA} \quad q_{211A} \quad q_{212A} \quad \cdots \quad q_{21NA} \quad q_{2M1A} \quad q_{2M2A} \quad \cdots \quad q_{2MNA}\}^T,$$
$$m = 0,1,2,\ldots,M, \quad n = 1,2,\ldots,N \tag{24}$$

We chose to use only the cosine DOF due to the problem's axisymmetric nature which yields doublet modes [32,33] (i.e., for each natural frequency where $m>0$ there is a cosine mode and a similar mode shifted in space by $\pi/2$, the sine mode). With this information, the stiffness and mass matrices can be computed (see Appendix B). It is visible from the topology of the matrices that the coupling between the openings is indeed only through the gas with no contribution from the fluid. Notably, only the axisymmetric functions of each opening ($m=0$) contribute to the coupling.

The potential given in Eq.(22) influences only the mass matrix (see Appendix B). Once the elements of **β** are calculated, as described in the following section, the stiffness and mass matrix can be computed. The generalized eigenproblem is then solved to compute the natural frequencies and the normal modes [28].

## 3. Natural frequencies and vibration modes – intermediate case

For the general case where the bubble lies on a finite plane, an additional parameter per each interface is required. To estimate these parameters, we devised an optimization problem that minimizes the error between the analytically and numerically obtained solutions as a function of $\boldsymbol{\beta}$. From the analytical solution, we can obtain the natural frequencies, normal vibration modes, and potential fields. Moreover, because a potential flow is assumed, the analytical pressure field can be computed as follows:

$$p_i^{An} = -j\omega\phi_i. \tag{25}$$

We assume a lossless harmonic problem; therefore, the pressure field can be computed numerically using a BEM simulation. The input to the simulation is the interface velocity, which is dictated by the analytically computed vibration modes and natural frequencies, and the output is the pressure field $p_i^{BEM}$. The intermediate case can be resolved by minimizing the error between the two fields:

$$\Delta = \sum_i \sqrt{\frac{\sum\left(p_i^{An}(\mathbf{x}) - p_i^{BEM}(\mathbf{x})\right)^2}{\sum\left(p_i^{BEM}(\mathbf{x})\right)^2}}. \tag{26}$$

We compute the natural frequencies and matching normal modes for the free, baffled, and intermediate cases for several geometries. We begin by analyzing a bubble with a single opening and demonstrate the impact of the geometry and optimization procedure on the simulation results. The values for the model parameters are chosen as follows:

$$\begin{aligned} &\sigma = 0.072\,\mathrm{N\,m^{-1}}, \quad \kappa = 1.4, \quad \rho_{water} = 998.24\,\mathrm{kg\,m^{-3}}, \quad c_{water} = 1481.4\,\mathrm{m\,s^{-1}}, \\ &p_0 = 101325\,\mathrm{Pa}, \quad h = 150\,\mathrm{\mu m}, \quad a = 20\,\mathrm{\mu m}, \quad R = 1.1. \end{aligned} \tag{27}$$

We reduced the computation time while maintaining a sufficient level of accuracy by truncating the basis function series and selecting $M=4$ and $N=3$, following convergence analysis. The truncated series provided accurate predictions of all the modes up to the third axisymmetric mode. The mode shapes of the baffled case are shown in Fig. 2 which are almost identical to the mode shapes of the free case.

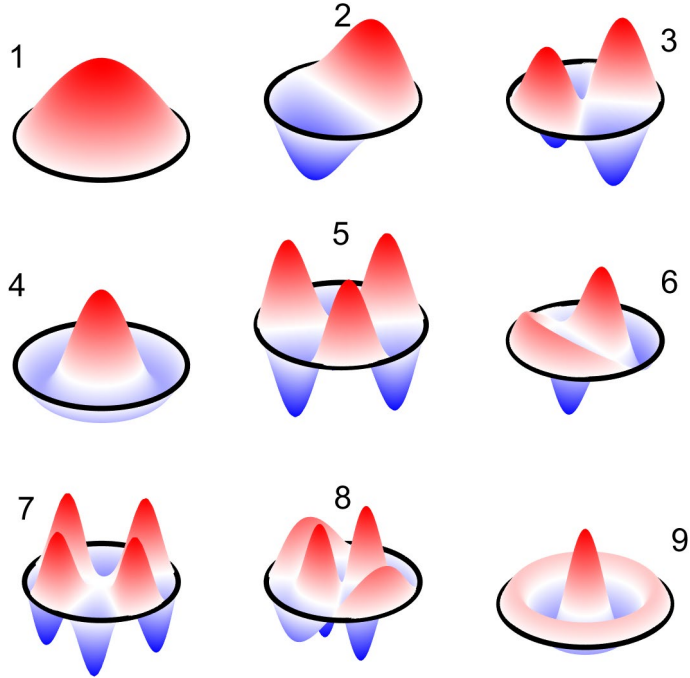

**Fig. 2.** The first nine analytically computed normal vibration modes of the baffled bubble. Modes 1, 4, and 9 are axisymmetric, while the rest are not. For each non-axisymmetric mode there should be a similar mode shifted in space by $\pi/2$ having the same natural frequency.

We are interested in finding the solution for the intermediate case when R is finite. To this end, the optimization problem that was described in Section 3 is solved using MATLAB's built-in functions from the Optimization and Global Optimization Toolboxes. The acoustic field is simulated using OpenBEM [34] according to the selected mode(s), where the input to each simulation is the interface velocity obtained from the modal analysis. By minimizing the error, the optimal value for **β** is calculated. In general, the obtained results show that the optimization scheme leads to an accurate estimation of the pressure fields by the analytical model. The natural frequencies for the three different cases are given in Table 1. The natural frequencies were computed using two optimization schemes. In the first, each mode was estimated separately (i.e., each mode yielded a different value for β). In the second scheme, all the considered modes were computed together (i.e., yielding a single value for β). The results highlight the importance of considering the geometry near the opening of the microbubble. Although the mode shapes are practically identical, the relative error between the computed natural frequencies for the extreme cases can be higher than 33% for the first mode. The relative error reduces below 10% for higher modes, nevertheless, the difference remains more than 10 kHz. Notably, the intermediate solution lies in between the two extreme cases, and as $R$ increases, the solution approaches the baffled solution as expected. The latter can be rationalized as follows; larger $R$ increases the flow resistance, thus increasing the inertia, which results in lower natural frequencies. We can conclude that the optimization scheme (single mode versus all modes) has a negligible effect on the results, as the maximum obtained relative error is 1.32%. In practice, the uncertainty in the values of the parameters is expected to lead to larger errors.

**Table 1** Computed natural frequencies for a bubble with one opening

| Mode | | 1 | 3 | 4 | 8 | 9 |
|---|---|---|---|---|---|---|
| Baffled (kHz) | | 53.65 | 162.86 | 175.26 | 349.95 | 360.36 |
| Free (kHz) | | 71.51 | 175.87 | 198.31 | 369.04 | 385.44 |
| $R = 1.1$ | Single mode (kHz) | 57.33 | 163.85 | 178.57 | 351.65 | 363.48 |
|  | All modes (kHz) | 57.01 | 166.01 | 178.54 | 354.06 | 363.51 |
|  | R.E. (%) | 0.56 | 1.32 | 0.02 | 0.69 | 0.01 |
| $R = 2$ | Single mode (kHz) | 55.56 | 162.86 | 176.67 | 350.02 | 361.58 |
|  | All modes (kHz) | 55.10 | 164.27 | 176.64 | 351.76 | 361.68 |
|  | R.E. (%) | 0.83 | 0.87 | 0.02 | 0.50 | 0.03 |
| $R = 3$ | Single mode (kHz) | 54.90 | 162.86 | 176.14 | 349.95 | 361.11 |
|  | All modes (kHz) | 54.58 | 163.78 | 176.14 | 351.12 | 361.20 |
|  | R.E. (%) | 0.58 | 0.56 | 0.00 | 0.33 | 0.02 |

Following a similar procedure, the natural frequencies and vibration modes of an entrapped bubble with two openings were computed. The first nine vibration modes for the baffled case with the following parameters $a_1 = 1$, $a_2 = 0.6$ (see Eq.(27) and Fig. 1, the opening radii are 20 μm and 12 μm) are shown in Fig. 3. As for a bubble with a single opening, the mode shapes dependency on $R$ is negligible and they resemble, however, the natural frequencies vary considerably as shown in Table 2. For the selected parameters, both interfaces are deformed only in the first two modes, and at higher modes, only one is deformed. This characteristic can be used for frequency-selective actuation, where oscillations can be generated on both sides or at a single side. Each opening deformation shape can be correlated to the single opening bubble, and it is evident that the order for each opening is maintained (i.e., top opening modes 1, 3, 4, 6, 7, and 8 correlate to modes 1-6 in Fig. 2, and bottom opening modes 2, 5 and 9 correlate to modes 1-3 in Fig. 2).

**Table 2** Computed natural frequencies for a bubble with two openings

| Mode | | 1 | 2 | 3 | 5 | 6 |
|---|---|---|---|---|---|---|
| Baffled (kHz) | | 51.95 | 79.78 | 100.39 | 167.31 | 175.04 |
| Free (kHz) | | 69.42 | 96.97 | 113.34 | 188.91 | 198.35 |
| $R_1 = 1.1$ $R_2 = 1.1$ | Single mode (kHz) | 66.30 | 73.89 | 102.13 | 175.46 | 178.44 |
| $R_1 = 1.1$ $R_2 = 2$ | Single mode (kHz) | 64.18 | 73.90 | 102.09 | 179.61 | 178.39 |
| $R_1 = 1.1$ $R_2 = 3$ | Single mode (kHz) | 63.60 | 73.90 | 102.04 | 179.71 | 178.33 |

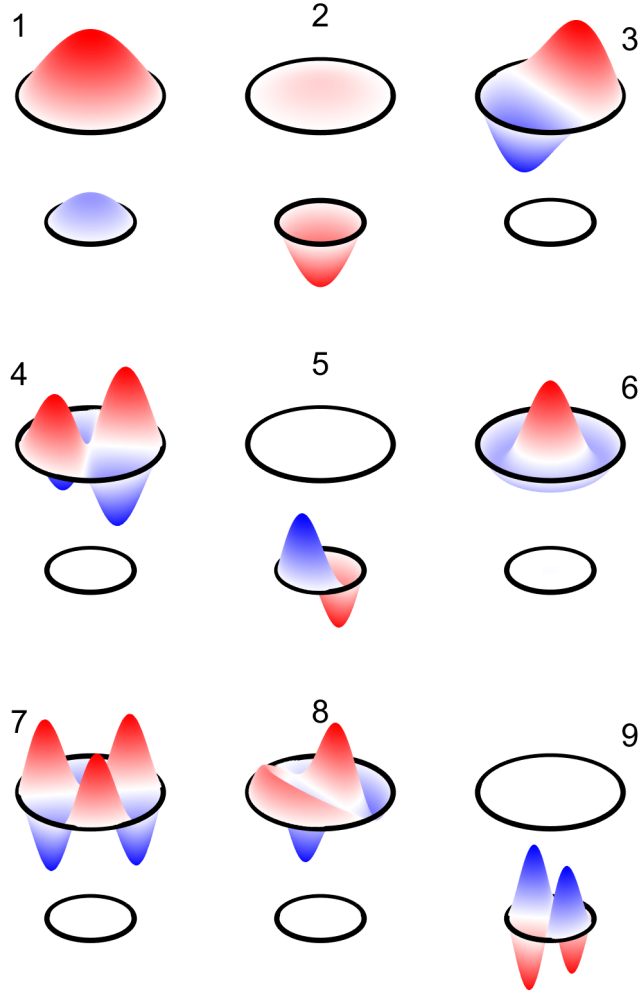

**Fig. 3.** The first nine analytically computed normal vibration modes assuming both interfaces are baffled, where the encapsulating structure is not shown for simplicity. The top opening is larger, and modes 1, 2, and 6 are axisymmetric, while the rest are not. For each non-axisymmetric there should be a similar mode shifted in space by $\pi/2$ having the same natural frequency. For modes 3 and higher, the deflection of one opening is negligible, and all the energy is concentrated in a single one.

When realizing this method, it is important to acknowledge the errors that may occur due to the discrepancies between the free and baffled bubble solutions, and related truncation errors. As mentioned in the introduction, commonly the first natural frequency is selected as the operational frequency. To compute it, the method assumes that in both extreme cases the mode shape is the same. However, this is not true for all geometries as can be seen in Fig. 4, which shows, for a single opening bubble, the two first natural frequencies of the extreme cases as a function of the bubble's radius. Beyond $a \approx 48.5$ μm for the baffled bubble and $a \approx 56$ μm for the free bubble, the mode shapes are different. Truncation of the basis function series may lead to similar results; therefore, it is important to perform a proper convergence analysis. Similar errors may occur for higher natural frequencies. This observation was previously reported and justified by Gelderblom *et al.*, [20].

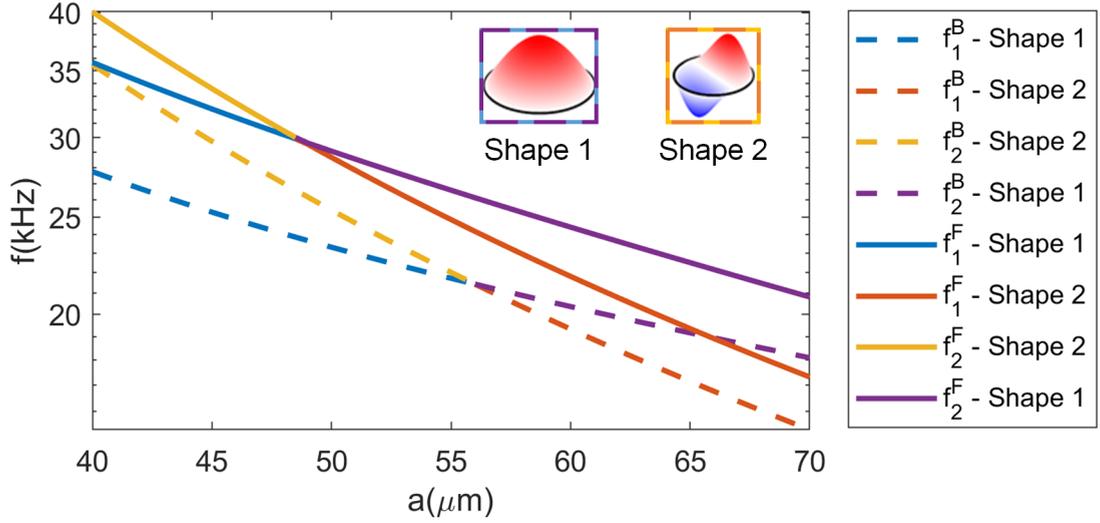

**Fig. 4.** The first two natural frequencies and matching vibrations modes shapes as a function of the bubble radius for the baffled (dashed lines) and free (continuous lines) cases. The two possible vibration mode shapes for both cases exchange their order for different values of the bubble radius.

### 4. Acoustic streaming and thrust

We employed linear harmonic analysis to derive the modal characteristics of the system. The generated thrust and AS are nonlinear phenomena; therefore, cannot be directly computed with a linear model. For the latter, thermoviscous losses and higher-order terms should be considered. The losses give rise to the AS phenomenon by attenuating acoustic waves resulting in a transfer of pseudo-momentum from the wave to the fluid [26]. For the microbubbles, the dominating attenuation occurs at the boundaries due to the existence of a viscous boundary layer. By considering higher order terms, the thrust and AS can be approximated by time-averaging the stress acting on the bubble and the velocity field accordingly. The equations governing the thermoacoustic fields are well established, and the fields are usually written as perturbation series [35]:

$$\begin{aligned} T &= T_0 + T_1 + T_2, \\ p &= p_0 + p_1 + p_2, \\ \mathbf{v} &= \mathbf{v}_1 + \mathbf{v}_2, \end{aligned} \tag{28}$$

where $T$, $p$, and $v$ denotes the temperature, pressure, and velocity fields, respectively. The subscript 0 represents the ambient conditions, where it is assumed that there is no mean flow. The subscript 1 and 2 represent the linear harmonic fields and the second-order nonlinear fields, respectively.

Due to the geometric complexity, a numerical approach is adopted here according to the work by Muller *et al.*,[26]. First, the harmonic acoustic thermoviscous problem is solved in COMSOL Multiphysics 5.5. Then, the obtained solution (i.e., $p_1$ and $\mathbf{v}_1$) is used to solve the second-order and

$\langle p_2 \rangle$ and velocity field $\langle \mathbf{v}_2 \rangle$, which are used to compute the streaming patterns and estimate the total force applied by the bubble (i.e., thrust) [29]:

$$\mathbf{F} = \int_{\partial \Omega} \left[ \langle \sigma_2 \rangle - \rho_{water} \langle \mathbf{v}_1 \mathbf{v}_1 \rangle \right] \cdot \mathbf{n} da, \tag{29}$$

the angled brackets represent the time-average operator, $\partial \Omega$ is a static surface surrounding the bubble, $\sigma_2$ is the second-order stress field, and $\mathbf{n}$ is a surface vector into the fluid.

*Sensitivity analysis*

To perform sensitivity analysis we analyze the axisymmetric vibration modes of a bubble with a single opening. The first three axisymmetric modes for $R = 1.1$ are shown in Fig. 2. The procedure involves the analytical computation of the optimal solution for each mode separately. Next, the velocity distribution and the natural frequency for each mode were used as inputs to COMSOL to solve the thermosviscous problem and compute the second-order time-averaged fields. The velocity fields along with the streamlines are shown in Fig. 5. The streamlines for the various cases resemble, however, as higher vibration modes are considered the main vortex ring (highlighted in red in Fig. 5) changes and approaches the bubble's surface. Additional vortexes, which are also highlighted in red, are formed and the microjets emanating from the bubble's center become narrower (i.e., the radial distance to the vortex ring becomes smaller). Furthermore, higher velocities are obtained for higher vibration modes.

The generated thrust, which is defined as the net force in the z-direction, was computed numerically according to Eq.(29). The thrust increases with the frequency, and the natural frequencies decrease with increasing values of $R$ and $a$. Therefore, to study the contribution of the mode shape alone (i.e., canceling the contribution of the frequency) to the thrust generation, we computed the stress for a maximum displacement of 1 μm (Fig. 6(a)) and maximum velocity of 0.1 m/s Fig. 6(b)). In the latter case, since the velocity is set, the excitation frequency cancels out. The computed thrust concerning $R$ and $a$ for the first three axisymmetric modes is shown in Fig. 6. For the fixed displacement case, higher modes generate more thrust and $R$ has a negligible effect. Surprisingly, larger bubbles (i.e., larger $a$) generate less thrust in this scenario. In the second scenario, where constant velocity was used, lower modes and larger bubbles generate more thrust, as expected, while the influence of $R$ remains negligible. The comparison between the two scenarios, highlights the importance of the frequency and velocity as argued in the introduction (i.e., natural versus resonance frequency). The thrust is highly dependent on the frequency, therefore lower modes and larger bubbles which have lower natural frequencies generate less thrust for the same displacement amplitude.

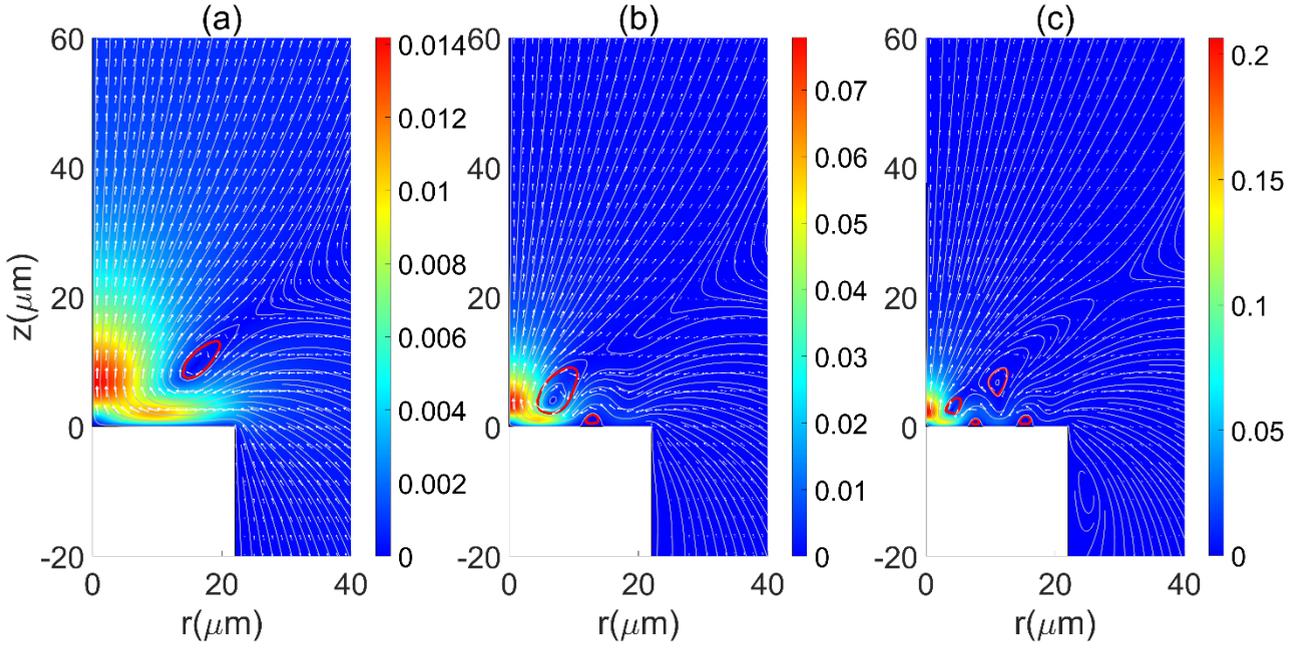

**Fig. 5.** The acoustic streaming patterns that are generated by the first three axisymmetric modes. The color map indicates the magnitude of the particle velocity $|\langle v_2 \rangle|$ m s$^{-1}$, arrows indicate the direction, streamlines are shown in gray and some vortexes are highlighted in red. Panels (a), (b), and (c) correlate to modes 1, 4, and 9. The maximum displacement of the interface was set to 1 μm.

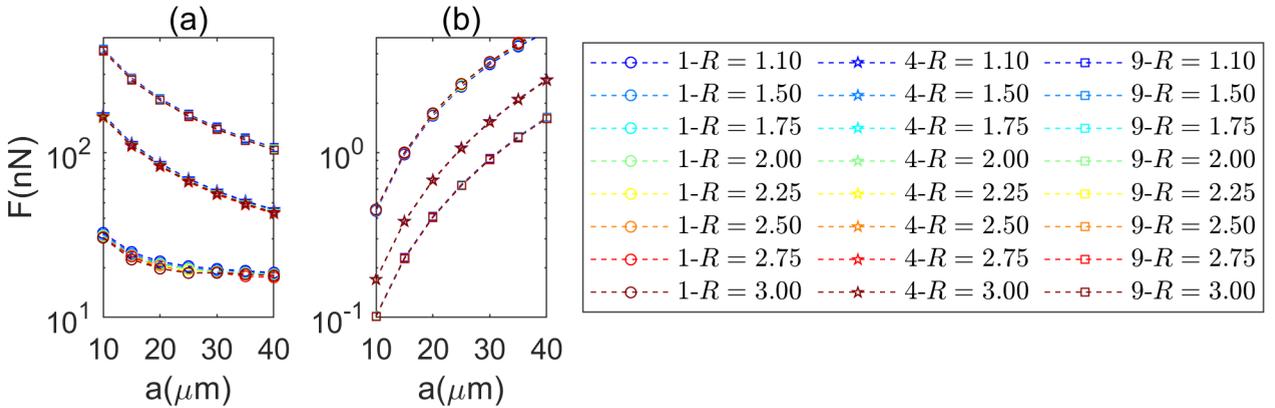

**Fig. 6.** The numerically computed thrust that is generated by a bubble with a single opening. The results are shown for the first three axisymmetric modes for various values of $R$, where in (a) a maximum amplitude of 1 μm was used, and in (b) a maximum velocity of 0.1 m s$^{-1}$ was used.

## 5. Encapsulated microbubble with two openings

In this section, we derive the governing equations of motion of an encapsulated bubble with two circular openings, as shown in Fig. 7(a). The capsule holding the bubble was 50 μm long, had an inner radius of 20 μm, an outer radius of 25 μm, and had two openings on both sides, with radii of 12 μm and 7 μm. We chose the dimensions such that the bubble can be treated as baffled, and that the first two natural frequencies are within the frequency band of the water immersion transducer (see Fig. S2(b) in the Supplementary Material). Moreover, to stabilize the bubble we use saline water (25% salinity) [36,37]. The following parameters were used for the analysis.

$$\begin{aligned}
&\sigma = 0.079\,\mathrm{N\,m^{-1}}, \quad \kappa = 1.4, \quad \rho_{\mathrm{water}} = 1147.8\,\mathrm{kg\,m^{-3}}, \quad c_{\mathrm{water}} = 1776.1\,\mathrm{m\,s^{-1}}, \\
&p_0 = 101325\,\mathrm{Pa}, \quad V_0 = 6.2832 \times 10^4\,\mathrm{\mu m^3}, \quad a = 12\,\mathrm{\mu m}, \\
&\alpha_1 = 1, \quad\quad\quad\quad \alpha_2 = 7/12, \quad\quad R_1 = R_2 \to \infty.
\end{aligned} \quad (30)$$

According to the chosen parameters, we computed the natural frequencies and vibration modes. The first two natural frequencies are 97.36 kHz and 153.46 kHz, and the matching vibration modes resemble the ones in Fig. 3. At these frequencies, the acoustic wavelength is much longer than the openings' diameter, even if we consider much higher frequencies (e.g., at 1 MHz, $\lambda \approx 1.5$ mm). As a result, we can treat the pressure wave impinging the bubble as uniform (see Supplementary Material) and we can easily project it on the basis functions and mode shapes through an inner product. To better approximate the system, losses which were overlooked thus far should be added to the model via modal damping. Estimating the modal damping coefficients is a challenging task because there are many loss mechanisms, such as acoustic radiation to infinity, losses in the gas and fluid, and heat conduction [8,38–40]. In general, the loss mechanisms are not linear as the bubble's dynamics [41]. However, considering the results obtained by Gelderblom *et al.*, [20] and the parameters that are given in Eq.(30) we assume uniform modal damping of 20%. Now, the governing equations in modal coordinates [28] **η** are given by

$$\ddot{\boldsymbol{\eta}} + \boldsymbol{\Gamma}\dot{\boldsymbol{\eta}} + \boldsymbol{\Lambda}\boldsymbol{\eta} = \boldsymbol{\Phi}^T\mathbf{Q} . \tag{31}$$

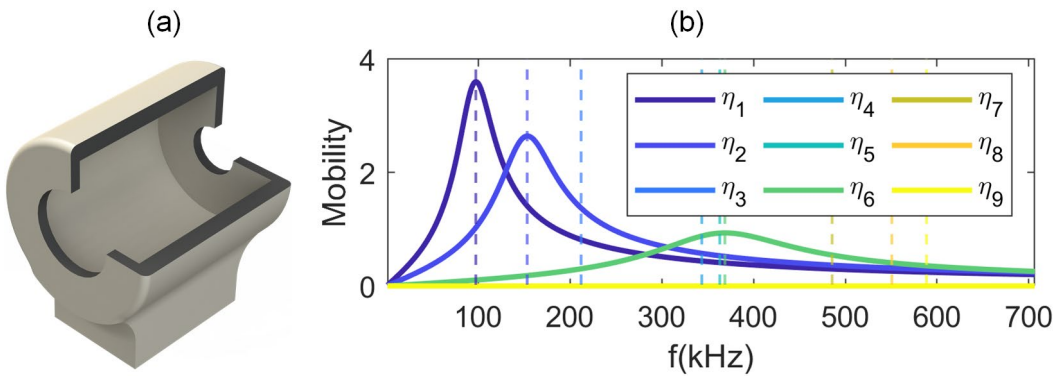

**Fig. 7.** (a) Illustration of the capsule holding the bubble with a section cut for visualization. (b) Modal mobility plot, showing the first nine modes' response to a uniform pressure excitation, and dashed lines highlight the matching natural frequencies.

Here **I** is the identity matrix, **Γ** is the modal damping matrix, **Λ** is the modal stiffness matrix, **Φ** is the modal matrix whose columns are the normal modes and **Q** is the uniform pressure projection on the basis functions. **Γ** is a diagonal matrix whose main entries are $2\zeta_i\omega_i$, and **Λ** is a diagonal matrix whose main entries are $\omega_i^2$ where $\zeta_i$ and $\omega_i$ are the modal damping and natural frequency correlated to the $i^{th}$ mode. For an impinging pressure wave with an amplitude of 0.5 kPa, the modal mobility plot for the first nine modes is shown in Fig. 7(b). A maximum displacement amplitude of around 1.4 μm is obtained at the center of the bigger opening. Only the axisymmetric modes (see modes 1, 2, and 6 in Fig. 3) are excited, and the first mode obtains the largest mobility at the associated natural frequency. The following axisymmetric modes also obtain large mobility values near their associated natural frequencies.

*Experimental estimation of the natural frequencies*

The microstructures entrapping microbubbles with two openings were 3D printed using a two-photon polymerization technique (Nanoscribe Photonic Professional GT+, 3D laser writer), from a photopolymerizable polymer, IP-Dip, directly on a quartz glass substrate [44]. We closed a microfabricated Poly(dimethylsiloxane) (PDMS) microfluidic device around the structures to facilitate particle image velocimetry. Tracer particles (Polysciences - 1μm polystyrene microspheres) were injected into the microchannel to visualize the AS patterns, which were recorded by a high-speed camera (Phantom EVO640L). A surfactant (TWEEN 20) was added to the saline water to avoid the tracer particles' agglomeration. We fabricated only the capsule holding the bubble, as shown in Fig. 7(a). The microstructures were placed in a water tank atop an inverted microscope (Nikon Eclipse Ti-2-U). To excite the bubbles, a water immersion transducer (Ultran group, GPS100-D19) was used. The normalized spectrum of the transducer (see Fig. S2 in the Supplementary Material) was measured with a hydrophone (RP acoustics e.K. PVDF type s).

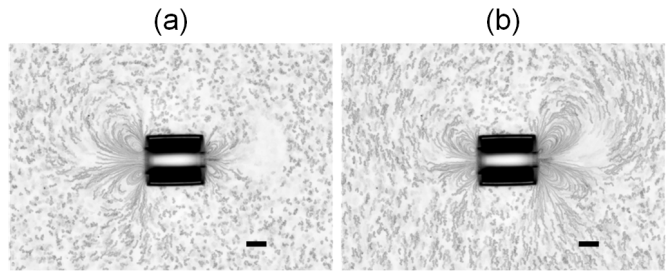

**Fig. 8.** The Acoustic streaming patterns that were generated by an entrapped bubble with two openings at the same pressure amplitude of 0.25 kPa, and two frequencies. (a) At 71.2 kHz the dominant AS is near the large opening on the left, and (b) at 124.1 kHz the dominant AS is near the small opening on the right. Scale bars, 20 μm.

To estimate the natural frequencies, we did a frequency sweep and observed that AS was generated in a wide range of frequency bands. Yet, at certain bands, the AS was more dominant around one opening or the other. While frequencies close to 71.2 kHz, which corresponds to the first natural frequency of the bubble, generated a significantly stronger flow outside the larger opening, at frequencies close to 124.1 kHz, which corresponds to the second natural frequency of the bubble, the dominant AS was recorded around the smaller opening (Fig. 8 and Video 1). The pressure amplitude was approximately 0.25 kPa at both frequencies. The discrepancy between the theoretical (97.36 kHz and 153.46 kHz) and empirical values of the natural frequencies may be due to various reasons, such as 3D printing inaccuracies, and errors in the fluid and gas parameters. Nevertheless, it is observed that there is a uniform shift of about 27 kHz between the experimental and theoretical results. Notably, we were able to obtain the desired behavior where we can selectively excite each mode.

*Qualitative estimation of the thrust*

After a successful estimation of the natural frequencies, we wanted to qualitatively estimate the generated thrust. Using Eq.(31), we computed both interface velocity distribution in response to a uniform pressure field of 0.5 kPa at different frequencies. Then, using COMSOL we computed the second-order stress and the total thrust with Eq.(29). The obtained results are shown in Fig. 9(a). It is noticeable that the force decreases and reaches its minimal value close to the first natural frequency. Then, as the frequency is increased it changes its sign and obtains its maximal value close to the second natural frequency. As the frequency is further increased, the thrust reduces and changes its sign again, and obtains a local minimum close to the sixth natural frequency. As the frequency is further increased, the thrust decays to zero. In general, the calculated force magnitude is in the same order as observed in previous works [42,43].

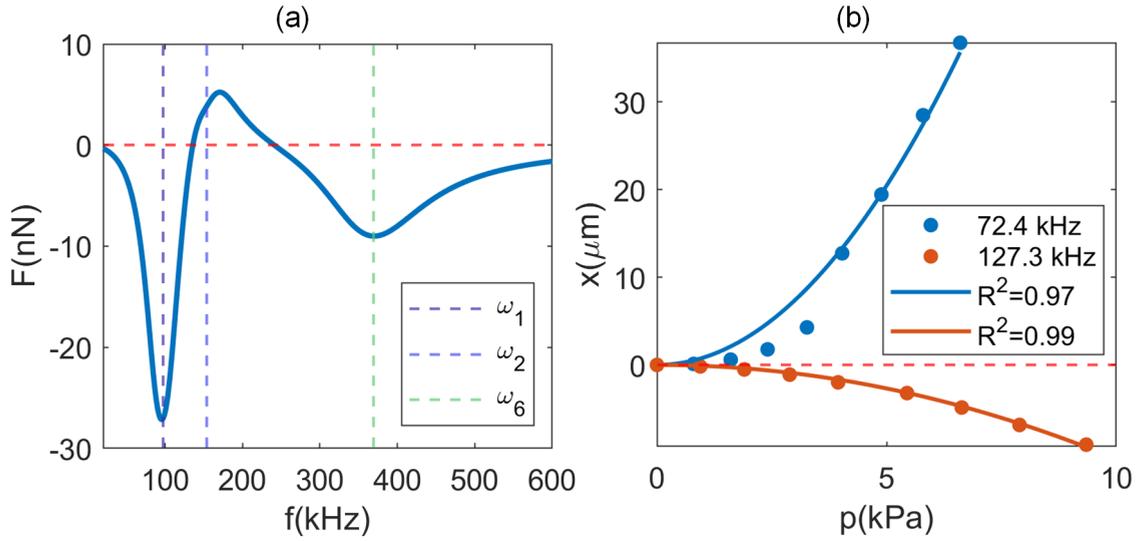

**Fig. 9.** (a) Numerically computed thrust versus the excitation frequency for a pressure wave of 0.5 kPa. The dashed lines highlight the first, second, and sixth natural frequencies. (b) The measured deflection at the beam's end at two frequencies versus the applied pressure amplitude, and the fitted curves.

Direct measurement of the thrust requires a force sensor with piconewton resolution. Moreover, the microstructures should be attached to the sensor, which is a challenging task. To circumvent the latter, we 3D printed flexible cantilever beams (56 μm long with a cross-section of 1.25 μm by 3.75 μm) from a biocompatible soft polymer, trimethylolpropane ethoxylate triacrylate (TPETA) [45]. The bubble actuator was attached to the tip of the cantilever, as shown in Fig. 10(a). We focused on the first two natural frequencies to reduce the complexity due to the rich spectrum of the transducer (Fig. S2). Following a similar procedure as in Section 5.1, the first two natural frequencies were estimated as 72.4 kHz and 172.3 kHz. Then, the bubble was excited at different input voltage amplitudes, while the deformation at the tip was tracked, and the pressure was measured using a hydrophone. The tip deflection was extracted using subpixel resolution image processing algorithm [46]. Plotting the displacement with respect to the measured pressure revealed that the deflection of the bubble was frequency and amplitude-dependent, Fig. 9(b). Snapshots from the motion of the device are shown in Fig. 10(b) and Fig. S5 along with the measured pressure. Even though the deflections were large and extend beyond the linear beam theory, we were able to obtain a good fit for a quadratic dependency of the deflection versus the pressure, according to theory [47]. The fitted models are $x=0.8183p^2$ ($R^2=0.9779$) at 72.4 kHz and $x=-0.1073p^2$ ($R^2=0.9965$) at 127.3 kHz. Following the theoretical results in Fig. 9(a), larger deflections were obtained close to the first rather than the second natural frequency. The experimentally measured displacement ratio is around 13% while the theoretical maximal force ratio is around 19%.

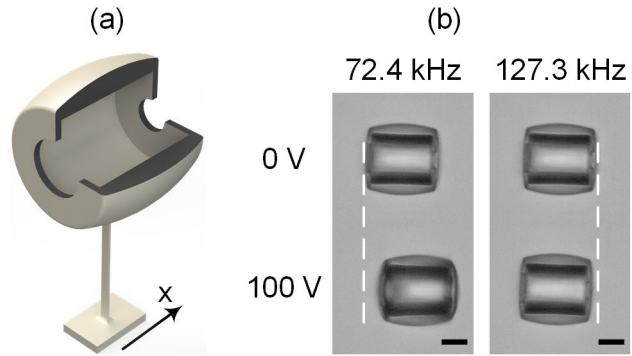

**Fig. 10**. (a) Illustration of the capsule holding the device attached to an upright cantilever beam with a section cut for visualization. (b) Bright-field microscopy images of the same device when it is not actuated (top images) and when it is actuated at two different frequencies, scale bar 20 μm.

## 6. Conclusions

We introduced an improved model to describe the dynamics of an arbitrarily shaped microbubble with multiple circular openings. The model can simulate more accurate geometries compared to the existing models by solving an optimization problem that couples acoustics and dynamics. Once the modal parameters are known, the system's response to external pressure can be computed. Then, the resulting AS and thrust can be computed numerically by a FEM simulation. The ability to estimate the natural frequencies and vibration modes was experimentally validated for a bubble with two openings, by visualizing the streaming patterns and qualitatively estimating the thrust.

The dynamical model comprises function series, which should be truncated for computational efficiency; however, a careful convergence analysis is required. Erroneous results may be obtained if too many elements are truncated. An additional source of errors arises if the two extreme cases (i.e., baffled and free) yield different orders of vibration modes, as discussed in Section 3. Using this model, the importance of the baffle region, $R$, was exemplified for several cases. The baffle region has a considerable influence on the natural frequencies, as shown in Table 1 and Table 2. Because microbubble-based devices are actuated with ultrasound transducers that usually have a narrow band, matching the natural frequencies of the device and the transducer is instrumental for higher efficiency.

The conventional wisdom suggests exciting the microbubbles at their resonance frequencies to achieve good performances. Nevertheless, optimal performances are obtained when bubbles are excited at the natural frequencies, at which the interface velocity, and not displacement, is maximal. This correlates well with the fundamentals of acoustics, where the field is driven by velocity rather than displacement. An additional advantage of natural frequencies is that they are inherent characteristics of the system and are independent of dissipation. The natural frequencies and vibration modes depend on multiple parameters. Here, we only focused on geometry. We studied the vibration modes of bubbles with one and two openings, and the generated streaming patterns and thrust. It was found that lower natural frequencies were manifested by larger and more baffled bubbles (i.e., larger values of $R$ and $a$). As a result, the generated thrust was also lower as thrust is highly dependent on the excitation frequency. On the other hand, it was also found that lower modes and larger bubbles are more efficient than higher modes. Therefore, selecting the optimal bubble size is not straightforward. An important yet often overlooked aspect of actuation is the excitation mechanism. It is unclear how the bubbles are excited, whether their structure is vibrated or the interface is excited by a pressure wave or both. Here, we briefly discussed the case where the interface is excited by a pressure wave and showed that if a uniform pressure is assumed only axisymmetric modes can be excited. Moreover, the latter also suggests that it should be easier and more efficient to excite mode shapes that resemble the first vibration mode of a membrane.

Loss mechanisms in general, and the modal damping in the suggested model has a crucial influence on the computed interface velocity that drives the acoustic phenomena. Estamating the modal damping is challenging as in other physical systems, and estimating it better can lead to an improved predictive model with which the thrust and streaming patterns are computed. In Section 4, we showed only streaming patterns that are generated by a bubble with a single opening oscillating in one of the axisymmetric modes. This methodology can be simply extended for 3D configurations, and general bubble osillations.

In a rationally designed bubble with two openings, the first two vibration modes resemble with an important difference that the openings oscillate in phase or anti-phase. Also, the deflection of each interface resembles the first mode of a bubble with a single opening, thus exciting both modes is feasible. Because each mode has a distinct frequency, they are frequency-selective, and the streaming and thrust can be controlled by the excitation frequency. The complexity of the problem highlights the contribution of the suggested simplified model as a design tool for the development of advanced acoustic micromachines and manipulators.

**Acknowledgment**

This work was partially supported by the European Research Council (ERC) under the European Union's Horizon 2020 research and innovation program (Grant agreement No. 714609). We would like to thank Theodoros Tsoulos for his assistance in the design of the optical setup, Guillermo Villanueva for generously providing the vibrometer, Izhak Bucher for fruitful discussions, EPFL Center of MicroNano Technology (CMi) staff for their technical assistance, and members of the Microbiorobotic Systems (MICROBS) Laboratory for their intellectual feedback.

**Appendix A**

The detailed equations of the potential and kinetic energies, and the relation between $a_{imn}$, $b_{imn}$, and $q_{imnA}$ and $q_{imnB}$ for the free and baffled bubbles are given below.

For the free bubble, the potential energy is

$$\mathcal{E}_p = \sum_i \left\{ \frac{\pi}{2} \sum_{m=1}^{\infty} \sum_{n=1}^{\infty} \sum_{q=1}^{\infty} q_{imnA} q_{imqA} m^2 F_{imnq} + \frac{\pi}{4\alpha_i^2} \sum_{q=1}^{\infty} \sum_{n=1}^{\infty} q_{i0nA} q_{i0qA} j_{0q} j_{0n} G_{i0nq} + \right.$$

$$\left. \frac{\pi}{8\alpha_i^2} \sum_{m=1}^{\infty} \sum_{n=1}^{\infty} \sum_{q=1}^{\infty} \left( q_{imnA} q_{imqA} + q_{imnB} q_{imqB} \right) j_{mn} j_{mq} G_{imnq} \right\} +$$

$$\kappa \lambda p_0 2\pi^2 \sum_i \sum_j \alpha_i^2 \alpha_j^2 \sum_{n=1}^{\infty} \sum_{q=1}^{\infty} q_{i0nA} q_{j0qA} \frac{J_1(j_{0n}) J_1(j_{0q})}{j_{0n} j_{0q}}, \quad (A1)$$

$$F_{imnq} = \int_0^{\alpha_i} J_m\left(j_{mq}\frac{r_i}{\alpha_i}\right) J_m\left(j_{mn}\frac{r_i}{\alpha_i}\right) \frac{1}{r_i} dr_i,$$

$$G_{imnq} = \int_0^{\alpha_i} \left[ J_{m-1}\left(j_{mn}\frac{r_i}{\alpha_i}\right) - J_{m+1}\left(j_{mn}\frac{r_i}{\alpha_i}\right) \right] \left[ J_{m-1}\left(j_{mq}\frac{r_i}{\alpha_i}\right) - J_{m+1}\left(j_{mq}\frac{r_i}{\alpha_i}\right) \right] r_i dr_i,$$

the relations between $a_{imn}$, $b_{imn}$ and $q_{imnA}$ and $q_{imnB}$ is

$$\dot{a}_{imn} = \dot{q}_{imnA}, \quad \dot{b}_{imn} = \dot{q}_{imnB}, \quad (A2)$$

and the kinetic energy is

$$\mathcal{E}_k^{(F)} = \frac{\pi}{2} \sum_i \sum_{n=1}^{\infty} \frac{\alpha_i^2 J_1^2(j_{0n})}{j_{0n}} \dot{q}_{i0nA}^2 + \frac{\pi}{4} \sum_i \sum_{m=1}^{\infty} \sum_{n=1}^{\infty} \frac{\alpha_i^2 J_{m+1}^2(j_{mn})}{j_{mn}} \left[ \dot{q}_{imnA}^2 + \dot{q}_{imnB}^2 \right]. \quad (A3)$$

For the baffled bubble, the relations between $a_{imn}$, $b_{imn}$ and $q_{imnA}$ and $q_{imnB}$ is

$$\dot{a}_{imn} \Phi_{im}(h) = \sum_{n=1}^{\infty} \frac{J_{m-1}(j_{mn}) J_m(h) j_{mn}}{h^2 - j_{mn}^2} \dot{q}_{imnA}, \quad \dot{b}_{imn} \Phi_{im}(h) = \sum_{n=1}^{\infty} \frac{J_{m-1}(j_{mn}) J_m(h) j_{mn}}{h^2 - j_{mn}^2} \dot{q}_{imnB}, \quad (A4)$$

and the kinetic energy is

$$\mathcal{E}_k^{(B)} = \pi\alpha_i^2 \sum_i \sum_{n=1}^{\infty} \sum_{q=1}^{\infty} \partial \dot{q}_{i0nA} \partial \dot{q}_{i0qA} J_{-1}(j_{0q}) J_{-1}(j_{0n}) j_{0n} j_{0q} f_{0nq}$$

$$+ \frac{\pi\alpha_i^2}{2} \sum_i \sum_{m=1}^{\infty} \sum_{n=1}^{\infty} \sum_{q=1}^{\infty} \left( \dot{q}_{imnA} \dot{q}_{imqA} + \dot{q}_{imnB} \dot{q}_{imqB} \right) J_{m-1}(j_{mn}) J_{m-1}(j_{mq}) j_{mn} j_{mq} f_{mnq}, \tag{A5}$$

$$f_{mnq} = \int_0^{\infty} \frac{J_m^2(k)}{\left(k^2 - j_{mn}^2\right)\left(k^2 - j_{mq}^2\right)} dk.$$

## Appendix B

According to the DOF order defined in Eq.(24), the elements of the mass and stiffness matrices can be computed. The stiffness matrix is the same for both cases and is given by

$$\mathbf{K} = \begin{pmatrix} \mathbf{K}_{10} & & \mathbf{K}_{120} & \\ & \mathbf{K}_{1m} & & \\ \mathbf{K}_{120} & & \mathbf{K}_{20} & \\ & & & \mathbf{K}_{2m} \end{pmatrix},$$

$$\left(\mathbf{K}_{10}\right)_{qn} = \frac{\pi}{2\alpha_1^2} j_{0q} j_{0n} G_{10nq} + \kappa\lambda p_0 4\pi^2 \alpha_1^4 \frac{J_1(j_{0n}) J_1(j_{0q})}{j_{0n} j_{0q}},$$

$$\left(\mathbf{K}_{1m}\right)_{qn} = \pi m^2 F_{1mnq} + \frac{\pi}{4\alpha_1^2} j_{mn} j_{mq} G_{1mnq}, \tag{B1}$$

$$\left(\mathbf{K}_{20}\right)_{qn} = \frac{\pi}{2\alpha_2^2} j_{0q} j_{0n} G_{20nq} + \kappa\lambda p_0 4\pi^2 \alpha_2^4 \frac{J_1(j_{0n}) J_1(j_{0q})}{j_{0n} j_{0q}},$$

$$\left(\mathbf{K}_{2m}\right)_{qn} = \pi m^2 F_{2mnq} + \frac{\pi}{4\alpha_2^2} j_{mn} j_{mq} G_{2mnq},$$

$$\left(\mathbf{K}_{120}\right)_{qn} = \kappa\lambda p_0 4\pi^2 \alpha_1^2 \alpha_2^2 \frac{J_1(j_{0n}) J_1(j_{0q})}{j_{0n} j_{0q}},$$

The mass matrices for the free and baffled cases differ. For the free bubble, the mass matrix is

$$\mathbf{M}^{(F)} = \begin{pmatrix} \mathbf{M}_{10}^{(F)} & & & \\ & \mathbf{M}_{1m}^{(F)} & & \\ \hline & & \mathbf{M}_{20}^{(F)} & \\ & & & \mathbf{M}_{2m}^{(F)} \end{pmatrix}, \quad \begin{aligned} \left(\mathbf{M}_{10}^{(F)}\right)_{nn} &= \frac{\pi \alpha_1^2 J_1^2(j_{0n})}{j_{0n}}, \\ \left(\mathbf{M}_{1m}^{(F)}\right)_{nn} &= \frac{\pi \alpha_1^2 J_{m+1}^2(j_{mn})}{2 j_{mn}}, \\ \left(\mathbf{M}_{20}^{(F)}\right)_{nn} &= \frac{\pi \alpha_2^2 J_1^2(j_{0n})}{j_{0n}}, \\ \left(\mathbf{M}_{2m}^{(F)}\right)_{nn} &= \frac{\pi \alpha_2^2 J_{m+1}^2(j_{mn})}{2 j_{mn}}, \end{aligned} \quad (B2)$$

For the baffled bubble, the mass matrix is

$$\mathbf{M}^{(B)} = \begin{pmatrix} \mathbf{M}_{10}^{(B)} & & & \\ & \mathbf{M}_{1m}^{(B)} & & \\ \hline & & \mathbf{M}_{20}^{(B)} & \\ & & & \mathbf{M}_{2m}^{(B)} \end{pmatrix}, \quad \begin{aligned} \left(\mathbf{M}_{10}^{(B)}\right)_{nq} &= 2\pi \alpha_1^2 J_{-1}(j_{0q}) J_{-1}(j_{0n}) j_{0n} j_{0q} f_{0nq}, \\ \left(\mathbf{M}_{1m}^{(B)}\right)_{nq} &= \pi \alpha_1^2 J_{m-1}(j_{mn}) J_{m-1}(j_{mq}) j_{mn} j_{mq} f_{mnq}, \\ \left(\mathbf{M}_{20}^{(B)}\right)_{nq} &= 2\pi \alpha_2^2 J_{-1}(j_{0q}) J_{-1}(j_{0n}) j_{0n} j_{0q} f_{0nq}, \\ \left(\mathbf{M}_{2m}^{(B)}\right)_{nq} &= \pi \alpha_1^2 J_{m-1}(j_{mn}) J_{m-1}(j_{mq}) j_{mn} j_{mq} f_{mnq}. \end{aligned} \quad (B3)$$

The stiffness matrix of the intermediate case remains unchanged while the mass matrix is computed as follows.

$$\mathbf{M}^{(I)} = \begin{pmatrix} e^{-\beta_1(R_1-\alpha_1)}\mathbf{M}_{10}^{(F)} & & & \\ & e^{-\beta_1(R_1-\alpha_1)}\mathbf{M}_{1m}^{(F)} & & \\ \hline & & e^{-\beta_2(R_2-\alpha_2)}\mathbf{M}_{20}^{(F)} & \\ & & & e^{-\beta_2(R_2-\alpha_2)}\mathbf{M}_{2m}^{(F)} \end{pmatrix}$$

$$+ \begin{pmatrix} \left(1-e^{-\beta_1(R_1-\alpha_1)}\right)\mathbf{M}_{10}^{(B)} & & & \\ & \left(1-e^{-\beta_1(R_1-\alpha_1)}\right)\mathbf{M}_{1m}^{(B)} & & \\ \hline & & \left(1-e^{-\beta_2(R_2-\alpha_2)}\right)\mathbf{M}_{20}^{(B)} & \\ & & & \left(1-e^{-\beta_2(R_2-\alpha_2)}\right)\mathbf{M}_{2m}^{(B)} \end{pmatrix}. \quad (B4)$$

# References

[1] Y. Li, X. Liu, Q. Huang, A.T. Ohta, T. Arai, Bubbles in microfluidics: an all-purpose tool for micromanipulation, Lab Chip. (2021). https://doi.org/10.1039/d0lc01173h.

[2] D. Ahmed, A. Ozcelik, N. Bojanala, N. Nama, A. Upadhyay, Y. Chen, W. Hanna-Rose, T.J. Huang, Rotational manipulation of single cells and organisms using acoustic waves, Nat. Commun. 7 (2016) 1–11. https://doi.org/10.1038/ncomms11085.

[3] A. Aghakhani, O. Yasa, P. Wrede, M. Sitti, Acoustically powered surface-slipping mobile microrobots, Proc. Natl. Acad. Sci. U. S. A. 117 (2020) 3469–3477. https://doi.org/10.1073/pnas.1920099117.

[4] L. Ren, N. Nama, J.M. McNeill, F. Soto, Z. Yan, W. Liu, W. Wang, J. Wang, T.E. Mallouk, 3D steerable, acoustically powered microswimmers for single-particle manipulation, Sci. Adv. 5 (2019) eaax3084. https://doi.org/10.1126/sciadv.aax3084.

[5] D. Ahmed, M. Lu, A. Nourhani, P.E. Lammert, Z. Stratton, H.S. Muddana, V.H. Crespi, T.J. Huang, Selectively manipulable acoustic-powered microswimmers, Sci. Rep. 5 (2015). https://doi.org/10.1038/srep09744.

[6] S. Boluriaan, P.J. Morris, Acoustic Streaming: From Rayleigh to Today, Int. J. Aeroacoustics. 2 (2003) 255–292. https://doi.org/10.1260/147547203322986142.

[7] N. Riley, Steady streaming, Annu. Rev. Fluid Mech. 33 (2001) 43–65. https://doi.org/10.1146/annurev.fluid.33.1.43.

[8] T. Leighton, The Acoustic Bubble, Academic Press, 2012.

[9] N. Bertin, T.A. Spelman, O. Stephan, L. Gredy, M. Bouriau, E. Lauga, P. Marmottant, Propulsion of Bubble-Based Acoustic Microswimmers, Phys. Rev. Appl. 4 (2015) 064012. https://doi.org/10.1103/PhysRevApplied.4.064012.

[10] T.A. Spelman, O. Stephan, P. Marmottant, Multi-directional bubble generated streaming flows, Ultrasonics. 102 (2020) 106054. https://doi.org/10.1016/j.ultras.2019.106054.

[11] C. Wang, B. Rallabandi, S. Hilgenfeldt, Frequency dependence and frequency control of microbubble streaming flows, Phys. Fluids. 25 (2013) 022002. https://doi.org/10.1063/1.4790803.

[12] I. Turmo De Arcos Gonzalez, Forcing microbubbles in microfluidics, Universidad de Sevilla, 2019.

[13] T. Luo, M. Wu, Biologically Inspired Micro-Robotic Swimmers Remotely Controlled by Ultrasound Waves, Lab Chip. (2021). https://doi.org/10.1039/d1lc00575h.

[14] A. Dolev, M. Kaynak, M.S. Sakar, On-Board Mechanical Control Systems for Untethered Microrobots, Adv. Intell. Syst. (2021) 2000233. https://doi.org/10.1002/aisy.202000233.

[15] G. Regnault, C. Mauger, P. Blanc-Benon, A.A. Doinikov, C. Inserra, Signatures of microstreaming patterns induced by non-spherically oscillating bubbles, J. Acoust. Soc. Am. 150 (2021) 1188–1197. https://doi.org/10.1121/10.0005821.

[16] M. Versluis, S.M. Van Der Meer, D. Lohse, P. Palanchon, D. Goertz, C.T. Chin, N. De Jong, Microbubble surface modes, in: Proc. - IEEE Ultrason. Symp., 2004: pp. 207–209. https://doi.org/10.1109/ultsym.2004.1417703.

[17] D.B. Khismatullin, A. Nadim, Radial oscillations of encapsulated microbubbles in viscoelastic

liquids, Phys. Fluids. 14 (2002) 3534–3557. https://doi.org/10.1063/1.1503353.

[18] J. Loughran, R.J. Eckersley, M.-X. Tang, Modeling non-spherical oscillations and stability of acoustically driven shelled microbubbles, J. Acoust. Soc. Am. 131 (2012) 4349–4357. https://doi.org/10.1121/1.4707479.

[19] D.L. Miller, W.L. Nyborg, Theoretical investigation of the response of gas-filled micropores and cavitation nuclei to ultrasound, J. Acoust. Soc. Am. 73 (1983) 1537–1544. https://doi.org/10.1121/1.389415.

[20] H. Gelderblom, A.G. Zijlstra, L. van Wijngaarden, A. Prosperetti, Oscillations of a gas pocket on a liquid-covered solid surface, Phys. Fluids. 24 (2012) 122101. https://doi.org/10.1063/1.4769179.

[21] D. Gritsenko, Y. Lin, V. Hovorka, Z. Zhang, A. Ahmadianyazdi, J. Xu, Vibrational modes prediction for water-air bubbles trapped in circular microcavities, Phys. Fluids. 30 (2018) 082001. https://doi.org/10.1063/1.5037328.

[22] C. Chindam, N. Nama, M. Ian Lapsley, F. Costanzo, T. Jun Huang, Theory and experiment on resonant frequencies of liquid-air interfaces trapped in microfluidic devices, J. Appl. Phys. 114 (2013) 194503. https://doi.org/10.1063/1.4827425.

[23] O. Schnitzer, R. Brandão, E. Yariv, Acoustics of bubbles trapped in microgrooves: From isolated subwavelength resonators to superhydrophobic metasurfaces, Phys. Rev. B. 99 (2019) 195155. https://doi.org/10.1103/PhysRevB.99.195155.

[24] Artificial Micro-Devices: Armoured Microbubbles and a Magnetically Driven Cilium, University of Cambridge, 2017. https://aspace.repository.cam.ac.uk/handle/1810/269647 (accessed March 16, 2021).

[25] A. Dolev, S. Davis, I. Bucher, Noncontact Dynamic Oscillations of Acoustically Levitated Particles by Parametric Excitation, Phys. Rev. Appl. 12 (2019) 034031. https://doi.org/10.1103/PhysRevApplied.12.034031.

[26] P.B. Muller, R. Barnkob, M.J.H. Jensen, H. Bruus, A numerical study of microparticle acoustophoresis driven by acoustic radiation forces and streaming-induced drag forces, Lab Chip. 12 (2012) 4617–4627. https://doi.org/10.1039/c2lc40612h.

[27] B. Behdani, S. Monjezi, J. Zhang, C. Wang, J. Park, Direct numerical simulation of microbubble streaming in a microfluidic device: The effect of the bubble protrusion depth on the vortex pattern, Korean J. Chem. Eng. 37 (2020) 2117–2123. https://doi.org/10.1007/s11814-020-0656-5.

[28] D.J. Ewins, Modal testing: theory, practice, and application, Research Studies Press, 2000.

[29] J.T. Karlsen, H. Bruus, Forces acting on a small particle in an acoustical field in a thermoviscous fluid, Phys. Rev. E - Stat. Nonlinear, Soft Matter Phys. 92 (2015) 043010. https://doi.org/10.1103/PhysRevE.92.043010.

[30] H. Lamb, Hydrodynamics, 6th ed., Dover publications, 1945.

[31] S.S. Rao, Vibration of Continuous Systems, 2nd ed., John Wiley & Sons, 2019.

[32] J.Y. Chang, J.A. Wickert, Response of modulated doublet modes to travelling wave excitation, J. Sound Vib. 242 (2001) 69–83. https://doi.org/10.1006/jsvi.2000.3363.

[33] R. Gabai, D. Ilssar, R. Shaham, N. Cohen, I. Bucher, A rotational traveling wave based levitation device – Modelling, design, and control, Sensors Actuators, A Phys. 255 (2017) 34–

45. https://doi.org/10.1016/j.sna.2016.12.016.

[34] V. Cutanda Henriquez, P.M. Juhl, OpenBEM-an open source boundary element method software in acoustics, in: Internoise 2010, Lisbon, 2010: pp. 5796–5805. https://www.researchgate.net/publication/232075627 (accessed January 29, 2019).

[35] M.F. Hamilton, D.T. Blackstock, Nonlinear Acoustics, Academic Press, 1998.

[36] S.J. Kleis, L.A. Sanchez, Dependence of speed of sound on salinity and temperature in concentrated NaCl solutions, Sol. Energy. 45 (1990) 201–206. https://doi.org/10.1016/0038-092X(90)90087-S.

[37] W.M. Haynes, D.R. Lide, T.J. Bruno, CRC Handbook of Chemistry and Physics, 97th ed., CRC Press, Taylor & Francis Group, 2016.

[38] A.A. Doinikov, Bubble and Particle Dynamics in Acoustic Fields: Modern Trends and Applications, Research Signpost, 2005.

[39] Z.L. Tian, Y.L. Liu, A.M. Zhang, L. Tao, Energy dissipation of pulsating bubbles in compressible fluids using the Eulerian finite-element method, Ocean Eng. 196 (2020) 106714. https://doi.org/10.1016/J.OCEANENG.2019.106714.

[40] G. Zhou, A. Prosperetti, Modelling the thermal behaviour of gas bubbles, J. Fluid Mech. 901 (2020) 3. https://doi.org/10.1017/jfm.2020.645.

[41] W. Lauterborn, T. Kurz, Physics of bubble oscillations, Reports Prog. Phys. 73 (2010) 106501. https://doi.org/10.1088/0034-4885/73/10/106501.

[42] N. Bertin, T.A. Spelman, O. Stephan, L. Gredy, M. Bouriau, E. Lauga, P. Marmottant, Propulsion of Bubble-Based Acoustic Microswimmers, Phys. Rev. Appl. 4 (2015) 064012. https://doi.org/10.1103/PhysRevApplied.4.064012.

[43] J. Feng, J. Yuan, S.K. Cho, Micropropulsion by an acoustic bubble for navigating microfluidic spaces, Lab Chip. 15 (2015) 1554–1562. https://doi.org/10.1039/c4lc01266f.

[44] M. Kaynak, P. Dirix, M.S. Sakar, Addressable Acoustic Actuation of 3D Printed Soft Robotic Microsystems, Adv. Sci. (2020) 2001120. https://doi.org/10.1002/advs.202001120.

[45] M. Hippler, K. Weißenbruch, K. Richler, E.D. Lemma, M. Nakahata, B. Richter, C. Barner-Kowollik, Y. Takashima, A. Harada, E. Blasco, M. Wegener, M. Tanaka, M. Bastmeyer, Mechanical stimulation of single cells by reversible host-guest interactions in 3D microscaffolds, Sci. Adv. 6 (2020). https://doi.org/10.1126/SCIADV.ABC2648.

[46] M. Guizar-Sicairos, S.T. Thurman, J.R. Fienup, Efficient subpixel image registration algorithms., Opt. Lett. 33 (2008) 156–158. https://doi.org/10.1364/OL.33.000156.

[47] H. Bruus, Acoustofluidics 10: Scaling laws in acoustophoresis, Lab Chip. 12 (2012) 1578–1586. https://doi.org/10.1039/c2lc21261g.

# Supplementary material
## Dynamics of entrapped microbubbles with multiple openings

Amit Dolev, Murat Kaynak, Mahmut Selman Sakar

*Institute of Mechanical Engineering, École Polytechnique Fédérale de Lausanne, CH-1015 Lausanne, Switzerland*

### 1. Measuring the interface deformation of entrapped bubbles

We built an experimental platform comprising a water tank, a water immersion transducers (Ultran group, GPS100-D19), a hydrophone (RP acoustics e.K. PVDF type s), a high-speed camera (Phantom EVO640L), a laser vibrometer (Polytec CLV-2534-2), an inverted microscope (Nikon Eclipse Ti-2-U), peripheral electronics (pico Technology PicoScope 5243D, THORLABS HVA20) and microfabricated devices that entrap microbubbles of well-defined sizes. An illustration of the platform is shown in Fig. S1. The platform was designed to enable the direct measurement of the bubble interface velocity with a laser Doppler vibrometer through the microscope optics. The laser beam path is controlled by adjusting two 45-degree mirrors, and the focus is controlled by the focus-ring in the sensor's head. To reduce the refractive index discrepancies and enable enough light to reflect from the water-air interface, we used a high-magnification water immersion objective (Nikon CFI Apo NIR 40X W). We measured the interface velocity of entrapped bubbles that are faced downwards (i.e., towards the objective). We began by studying the frequency response of baffled bubbles with different radii. Using the hydrophone, we measured the pressure field produced by the water immersion transducers in a single location. We positioned the hydrophone as close to the bubble as possible while avoiding interference with the light path. Due to its dimensions, the hydrophone was positioned at the top right edge of the microfabricated device, as shown in the inset of Fig. S1. The pressure was used as the input and the velocity was used as the output during the characterization of the system frequency response. Using high precision microscope stage and mirrors, we were able to measure multiple locations on the same bubble, as shown in Fig. S3, Fig. S4, and Video S1.

To confirm that the recorded signals are indeed coming from the interface, we measured two arbitrary positions on the device, and the response of the glass slide. These measurements were performed by placing two reflective stickers to the glass surface. While measuring the four positions, the transducer was excited with an input peak-to-peak voltage of 2 V, while measuring the bubble, the transducer was excited with input peak-to-peak voltage of 0.3 V. It is evident from Fig. S2(a) that the measured velocity can be exclusively attributed to the interface. Fig. S2(b) shows the rich spectrum of the measured normalized pressure.

We microfabricated an elastomer device with 64 cylindrical cavities using soft lithography. Briefly, a master mold was fabricated through etching technique, and a Poly(dimethylsiloxane) (PDMS) replica was cast by heating the pre-polymer at 65 °C for 4 hrs. The heights of the cylinders were set at 150 µm and the radii were varied from 12.5 µm to 50 µm. The cavities were spaced 2 mm apart from each other to minimize their interaction [1,2]. The device was raised 0.5 mm from the glass slide to minimize the expected acoustic streaming (AS) interaction with the surface.

We performed a 10 s linear frequency sweep from 40 kHz to 160 kHz with low amplitude to ensure linear response. At this voltage levels, the pressure was too low to be measured by the hydrophone, therefore, we repeated the experiment with a considerably larger voltage amplitude, assuming a linear response of the hydrophone. The mobility response curves for three different bubbles at five different locations on each bubble are shown in Fig. S3. When analyzing the results, one should be cautious as zeros in the pressure frequency curve may artificially generate large mobility values (e.g., Fig. S3(a), 47.26 kHz and 60.13 kHz), which must not be interpreted as natural frequencies.

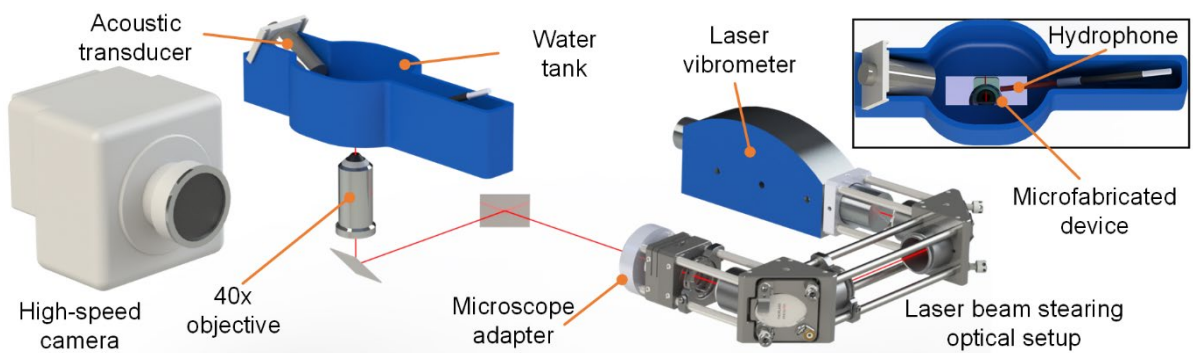

**Fig. S1**. Illustration of the experimental setup. The laser beam emitted by the vibrometer passes through two 45° adjustable mirrors that align and stir the beam. The optical setup is directly connected to an inverted microscope (not shown). The beam passes through two additional 45° mirrors inside the microscope before reaching the sample. The second internal mirror directs the beam towards the microscope objective, incoming light is focused on the sample, and the reflected light is collected by the vibrometer. High-speed imaging is performed using bright-field imaging. The water tank holding the specimen is shown in the inset. The tank holds an acoustic transducer, a hydrophone, and a glass slide that holds the microfabricated device with cavities trapping the bubbles (green).

We expect to see peaks in the mobility amplitude ratio diagrams near the natural frequencies for the three bubbles. The recorded spectrum, however, presented a multitude of peaks with varying amplitudes. In practice, each non-axisymmetric mode (i.e., modes 2, 3 and 5) represents doublet modes having slightly different natural frequencies, which may contribute to the complex spectrum. Despite these challenges, the first natural frequency can still be estimated form the mobility response of the smallest bubble (Fig. S3(a)). The natural frequency is expected to be 51.6 kHz from theoretical calculations, and the nearest peak in the recorded data is at 55.37 kHz. To estimate higher natural frequencies, the relative phase (i.e., relative to the phase of each bubble's center) could be used [3]. The data has shown that a phase separation occurs near the second theoretical natural frequency for all three bubbles (Fig. S3). The phase separation is an indication that an additional vibration mode is excited at that frequency. Theoretically, at each frequency, the instantaneous vibration shape can be described by a superposition of the normal vibration modes (i.e., standing waves); therefore, the phase shift between the points should be either 0 or $\pm\pi$, and should not change smoothly. The system is complex and the phase shift may be caused by the waves travelling on the bubble surfaces' as reported by Xu et al.,[4]. Furthermore, the pressure exciting the bubble, which cannot be measured with existing equipment, changes with the frequency and comprises traveling and standing waves. The observation of a smooth phase transition was reported by Wang et al.,[3]. When natural frequencies higher than the second are considered, the slope of some of the relative phases change near the theoretical values. For example, after the third natural frequency, the phases are relatively constant, and after the fourth, the phases of p3 and p4 change again. Additional observation is the relative amplitudes of the various points at the same frequency. For all three bubbles, at all frequencies as the measurement is further away from the center (p1<p2≈p4<p3≈p5), lower amplitudes are obtained in accordance with the first mode. This indicates that the first mode always dominates the response of the bubbles, which can be justified by the excitation mechanism. In the studied frequency range, the wavelength varies from approximately 0.9 cm to 3.75 cm, which is much larger than the radii of the bubbles. Hence, it is reasonable to assume a nearly planar pressure wave, whose projection on the first mode is maximal.

We used modal analysis to identify the instantaneous participation factor of the vibration modes. We measured 33 points on a bubble with a radius of 27.5 µm (Fig. S4, a snapshot of Video S1). The modal analysis did not reveal additional information, and the first mode dominated the response as expected.

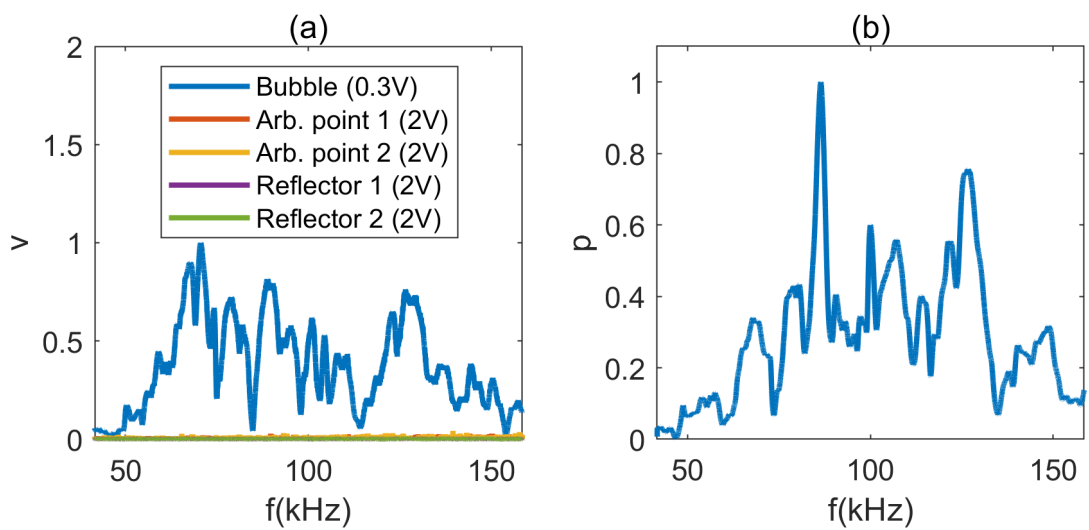

**Fig. S2.** (a) Normalized measured velocity at various locations. The nonzero measurement at the bubble's center when the input voltage to the transducer was 0.3 V peak-to-peak is shown blue, and the measurements at the other locations when the input was 2 V peak-to-peak collapse to the same line. (b) Normalized pressure measured in a single location close to the bubble.

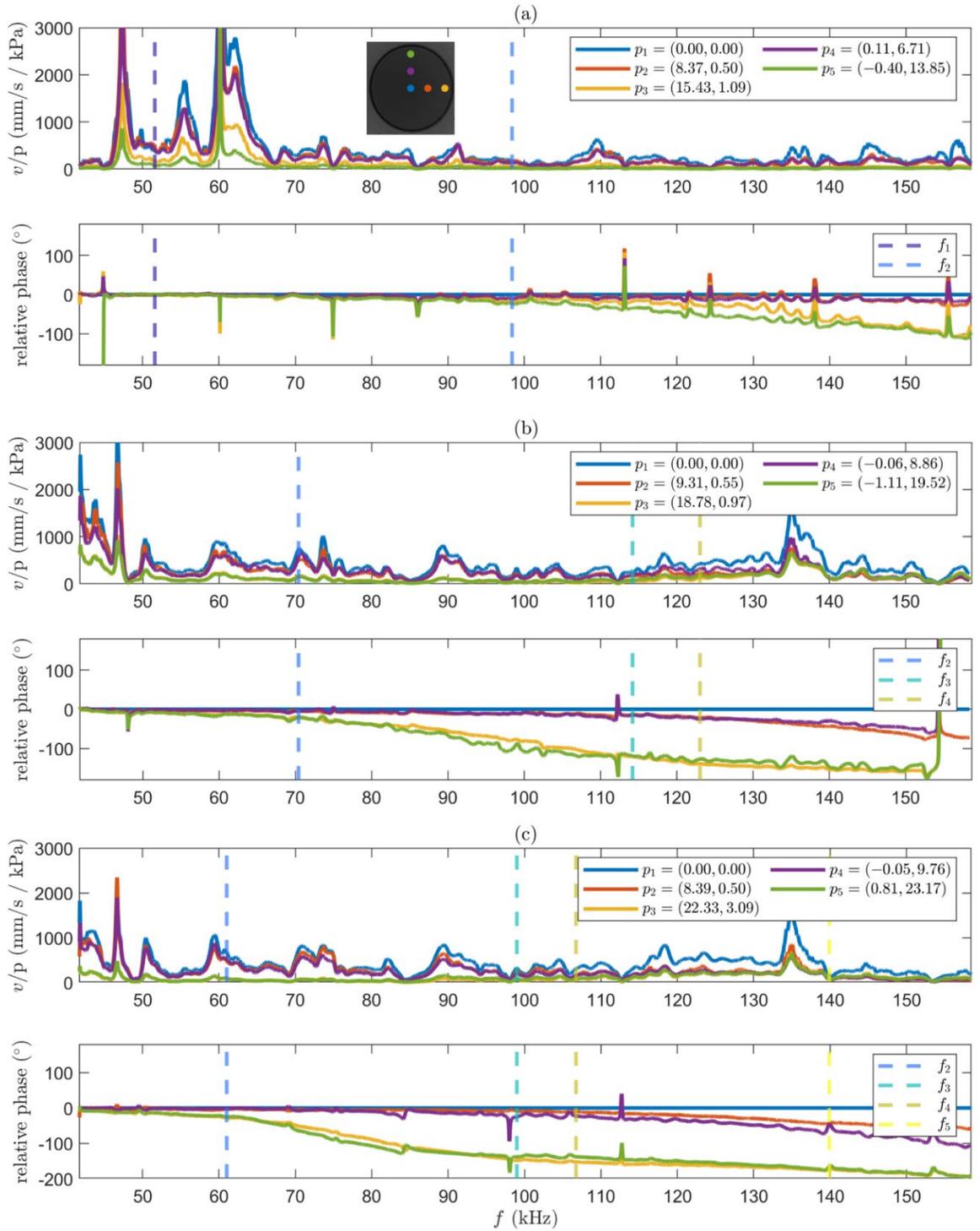

**Fig. S3**. The measured frequency responses of three different bubbles, (a) a = 20 μm, (b) a = 25 μm and (c) a = 27.5 μm. Each panel shows the mobility response curve as measured at five different locations on each bubble, which are color coded as shown in the inset of (a). The coordinates of each point are provided in the legends; units in μm. Vertical dashed lines highlight the theoretical estimated natural frequencies. The shown phase is relative to the phase of the point at the bubbles' center.

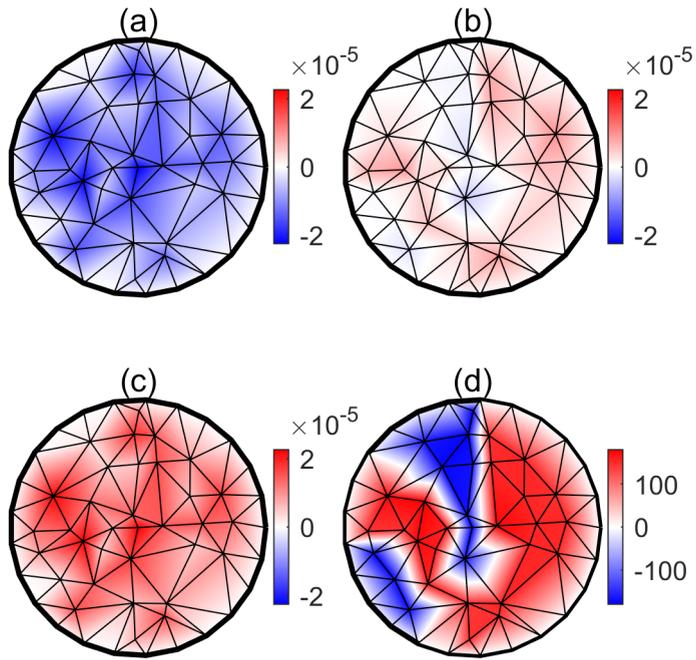

**Fig. S4**. A snapshot (69.561 kHz) from video 1 depicting the instantaneous identified real (a), imaginary (b), magnitude (c), and phase (d) of the frequency response as measured at 33 points on a bubble with radius of 27.5 μm.

## 2. Estimating the beam deflection

In Fig. S5, snapshots of the device are shown along with the input voltage to the transducer and resulting pressure. These snapshot were processed to extract the relative position of the device. The two last pictures (i.e., Amp = 180 and 200V) at 72.4 kHz could not be used to estimate the deflection and are therefore not provided in the manuscript.

| Amp (V) | 72.4 kHz p(kPa) | | 127.3 kHz p(kPa) | |
|---|---|---|---|---|
| 0 | 0 | 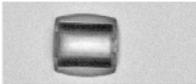 | 0 | 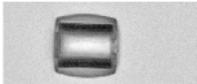 |
| 20 | 0.7969 | 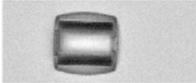 | 0.9437 | 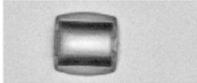 |
| 40 | 1.6051 | 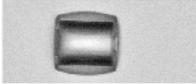 | 1.9017 | 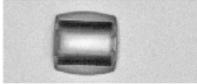 |
| 60 | 2.4086 | 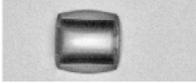 | 2.8895 | 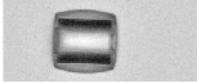 |
| 80 | 3.2651 | 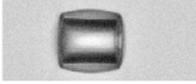 | 3.9437 | 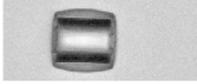 |
| 100 | 4.0327 | 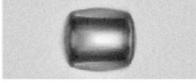 | 5.4445 | 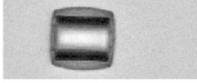 |
| 120 | 4.8869 | 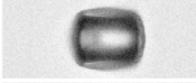 | 6.6357 | 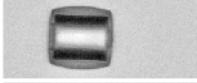 |
| 140 | 5.7941 | 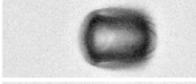 | 7.8939 | 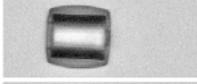 |
| 160 | 6.6016 | 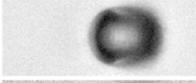 | 9.3462 | 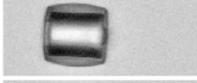 |
| 180 | 7.4476 | 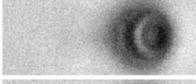 | 12.1773 | 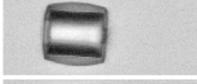 |
| 200 | 8.2493 | 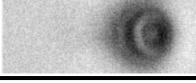 | 14.5395 | 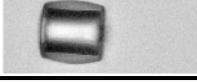 |

**Fig. S5.** Snapshots of the device deflection at two frequencies and different pressure amplitudes.

## References

[1]   D. Gritsenko, Y. Lin, V. Hovorka, Z. Zhang, A. Ahmadianyazdi, J. Xu, Vibrational modes prediction for water-air bubbles trapped in circular microcavities, Phys. Fluids. 30 (2018) 082001. https://doi.org/10.1063/1.5037328.

[2]   O. Schnitzer, R. Brandão, E. Yariv, Acoustics of bubbles trapped in microgrooves: From isolated subwavelength resonators to superhydrophobic metasurfaces, Phys. Rev. B. 99 (2019) 195155. https://doi.org/10.1103/PhysRevB.99.195155.

[3]   C. Wang, B. Rallabandi, S. Hilgenfeldt, Frequency dependence and frequency control of microbubble streaming flows, Phys. Fluids. 25 (2013). https://doi.org/10.1063/1.4790803.

[4]   J. Xu, D. Attinger, Acoustic excitation of superharmonic capillary waves on a meniscus in a planar microgeometry, Phys. Fluids. 19 (2007). https://doi.org/10.1063/1.2790968.